\documentclass[aps,preprint,showpacs,floatfix]{revtex4-1}
\usepackage{bm}
\usepackage{amsmath}
\usepackage{epsfig}
\usepackage{rotating}
\usepackage{graphicx}

\def\pr{\prime}
\def\be{\begin{equation}}
\def\lan{\left\langle}
\def\ran{\right\rangle}
\def\ee{\end{equation}}
\def\barr{\begin{array}}
\def\earr{\end{array}}

\def\l{\left}
\def\r{\right}
\def\dis{\displaystyle}
\def\ed{\end{document}}
\def\f{\frac}
\def\baa{{\mbox{\boldmath $\alpha$}}}

\def\co{{\cal O}}

\def\can{{\cal N}}

\def\dg{\dagger}

\def\la{\lambda}
\def\ed{\end{document}}
\begin{document}

\title{Multiple $SU(3)$ algebras in shell model and \\
interacting boson model}

\author{V.K.B. Kota}
\thanks{corresponding author, vkbkota@prl.res.in}
\affiliation{Physical Research Laboratory, Ahmedabad 380 009, 
India}
\author{R. Sahu}
\affiliation{National Institute of Science and Technology, Palur Hills, 
Berhampur-761008, Odisha, India}
\author{P.C. Srivastava}
\affiliation{Department of Physics, Indian Institute of Technology,Roorkee 
247 667, India}

\begin{abstract}

Rotational $SU(3)$ algebraic symmetry continues to generate new results in the
shell model (SM). Interestingly, it is possible to have multiple $SU(3)$
algebras for nucleons occupying an oscillator shell $\eta$. Several different
aspects of the multiple $SU(3)$ algebras are investigated using shell model and
also deformed shell model based on Hartree-Fock single particle states with
nucleons in $sdg$ orbits giving four $SU(3)$ algebras. Results show that one of
the $SU(3)$ algebra generates prolate shapes, one oblate shape and the other two
also generate prolate shape but one of them gives quiet small quadrupole moments
for low-lying levels. These are inferred by using the standard form for the
electric quadrupole transition operator and using quadrupole moments and $B(E2)$
values in the ground $K=0^+$ band in three different examples. Multiple  $SU(3)$
algebras extend to interacting boson model and using $sdg$IBM, the structure of
the four $SU(3)$ algebras in this model are studied by coherent state analysis 
and asymptotic formulas for  $E2$ matrix elements. The results from $sdg$IBM 
further support the conclusions from the $sdg$ shell model examples.

\end{abstract}


\maketitle

\section{Introduction}

Elliott has recognized way back in 1958 that shell model (SM) admits $SU(3)
\supset SO(3)$ algebra and this will generate rotational spectra in nuclei
starting with the interacting particle picture \cite{Ell-58a,Ell-58b}. 
Following this, $SU(3)$ algebra was developed in considerable detail by various
groups and this includes methods to obtain $SU(3)$ irreducible representations 
(irreps) and $SU(3)$ Wigner-Racah algebra with codes for calculating $SU(3)
\supset SO(3)$ and $SU(3) \supset SU(2) \times U(1)$ reduced Wigner
coefficients, $SU(3)$ Racah coefficients, $SU(3)$ coefficients of fractional
parentage and so on  \cite{KYbook,Ko-report,Dr-1,Verg,DrAk-1,DrAk-2,Dr-2,Dr-3}.
By mid 60's it was recognized that the $SU(3)$ symmetry is good for $1p$ and
$2s1d$ shell nuclei but due to the strong spin-orbit force it will be a badly
broken symmetry for $1p2f$ shell nuclei and beyond. Hecht, Draayer and others
later recognized \cite{Hecht-1,Hecht-2,Hecht-3,JPD-1,JPD-2} that for heavy
deformed nuclei, pseudo-$SU(3)$ based on pseudo spin and pseudo Nilsson orbits
will be a useful symmetry and it gave rise to many new results. Very recently, a
proxy-$SU(3)$ scheme by Bonatsos, Casten and others \cite{Bona-1,Bona-2,Bona-3}
has appeared within SM with definite prediction for prolate dominance over
oblate shape in heavy deformed nuclei. This $SU(3)$ model is currently being
investigated in more detail. In addition, in the multishell situation again
$SU(3)$ appears within the $Sp(6,R)$ model of Rowe and Rosensteel
\cite{RR-1,RR-2,RR-3} and this has given rise to the no-core-sympletic shell
model \cite{Krist-1,Krist-2}.  Going beyond SM, a major basis for the
interacting boson model (IBM) of atomic nuclei is that with $s$ and $d$ bosons
the spectrum generating algebra (SGA) is $U(6)$ and it has $SU(3)$ as a
subalgebra generating rotational spectrum \cite{Iac-76,Iac-87}. Similarly,
$sdg$IBM \cite{KY-rev,K-sdg}, $sdpf$IBM \cite{Long-1,Long-2} and also IBM-3 with
isospin ($T$) and IBM-4 with spin-isospin ($ST$) degrees of freedom 
\cite{Iac-87,KS} all contain $SU(3)$ symmetry generating rotational spectra. In
addition, in IBM-3 and IBM-4 models, $SU(3)$ also appears for isospin ($T$) and
spin-isospin ($ST$) degrees of freedom respectively. Similarly, for odd-A nuclei
we have $SU^{BF}(3) \times SU^F(2)$ symmetry in IBFM model with Nilsson
correspondence \cite{BK}. This extends to $SU(3)$ in IBFFM for odd-odd nuclei
\cite{KU-1,KU-2} and $SU(3)$ in IBF$^2$M for two quasi-particle excitations
\cite{KY-quasi}. With $SU(3)$ generating rotational spectra within both SM and
IBM, it is natural to look for new perspectives for $SU(3)$ symmetry in nuclei. 

One curious aspect of $SU(3)$ in nuclei is that in a given oscillator shell
$\eta$, there will be multiple $SU(3)$ algebras. Very early it is recognized
that in SM with $s$ and $d$ orbits there will be two $SU(3)$ algebras
\cite{parikh} but its consequences are not explored in any detail. Similarly, in
$sd$IBM there are two $SU(3)$ algebras \cite{Iac-87} and they are applied in
phase transition studies \cite{RMP}. Finally, it was also recognized that there
will be four $SU(3)$ algebras in $sdg$IBM \cite{K-sdg}. Except for the $sd$IBM,
properties of multiple $SU(3)$ algebras are not investigated in any detail in
the past. As we will show, for a given oscillator shell with major shell number
$\eta$, there will be $2^{\l[\f{\eta}{2}\r]}$ number of $SU(3)$ algebras where
$\l[\f{\eta}{2}\r]$ is the integer part of $\eta/2$. In the present paper,
following the recent investigation of multiple pairing algebras in SM and IBM
\cite{Ko-BJP}, several different aspects of multiple $SU(3)$'s in SM and IBM are
investigated. Now, we will give a preview.

In Section 2, multiple $SU(3)$ algebras in SM generated by angular momentum
operator $L^1_q$ and quadrupole moment operator $Q^2_q$ with different signs for
the $\ell \rightarrow \ell \pm 2$ matrix elements are identified and the matrix
elements for the corresponding $Q \cdot Q$ operators are given. Using these,
correlations between different $Q \cdot Q$ operators are studied. In Section 3,
Spectra and electric quadrupole ($E2$) properties of these algebras are studied
using shell model codes and also deformed shell model based on Hartree-Fock
single particle states (called DSM \cite{KS}). Used here are examples with 6
protons, 6 protons plus 2 neutrons and 6 protons plus 6 neutrons  systems. In
Section 4, results for multiple $SU(3)$ algebras in  IBM's (with no internal
degrees of freedom for the the bosons) are presented. Finally, Section 5 gives
conclusions.  

\section{Phase choice and Multiple $SU(3)$ algebras in shell model}

Let us consider the situation where valence nucleons in a nucleus occupying an
oscillator shell with major shell number $\eta$. With the spin-isospin degrees
of freedom for the nucleons, the spectrum generating algebra (SGA) is $U(4\can)$
and decomposing the space into orbital and spin-isospin ($ST$) parts, we have
$U(4\can) \supset U(\can) \times SU(4)$. Here, $\can=(\eta+1)(\eta+2)/2$ and
$SU(4)$ is the Wigner's spin-isospin $SU(4)$ algebra; see for example
\cite{Manan,Piet-su41,Piet-su42,octup,KS}. Also, for a given $\eta$, the the
single particle (sp) orbital angular momentum $\ell$ takes values $\ell= \eta$,
$\eta-2$, $\ldots$, $0$ or $1$. Note that, for nuclei with only valence protons
or neutrons $SU(4)$ changes to $SU(2)$ generating spin $S$. As Elliott has
established, the orbital $U(\can)$ algebra  admits $SU(3)$ subalgebra with
$U(\can) \supset SU(3) \supset SO(3)$ where $SO(3)$ generates orbital angular
momentum. The eight generators of $SU(3)$ are the orbital angular momentum
operators $L^1_q$ and quadrupole moment operators $Q^2_q$. In $LST$ coupling and
using fermion creation ($a^\dagger$) and annihilation ($a$) operators,
\be
L^1_q = 2 \dis\sum_\ell \dis\sqrt{\dis\frac{\ell(\ell+1)(2\ell+1)}{3}}
\l(a^\dagger_{\ell \f{1}{2} \f{1}{2}} \tilde{a}_{\ell \f{1}{2} \f{1}{2}}\r)^{
1,0,0}_q \;.
\label{eq.su31}
\ee
Note that $\tilde{a}_{\ell -m, \f{1}{2}-m_s, \f{1}{2}-m_t}=(-1)^{\ell-m +
\f{1}{2}-m_s +\f{1}{2} -m_t} a_{\ell m, \f{1}{2} m_s, \f{1}{2} m_t}$ where
$m_s$  and $m_t$ are the $S_z$ and $T_z$ quantum numbers for a single nucleon. 
Similarly, the quadrupole operator is
\be
Q^2_q = 2 \dis\sum_{\ell_f , \ell_i} \dis\frac{\lan \eta,\ell_f \mid\mid Q^2 
\mid\mid \eta,\ell_i\ran}{\dis\sqrt{5}} \l(a^\dagger_{\ell_f \f{1}{2} \f{1}{2}} 
\tilde{a} _{\ell_i \f{1}{2} \f{1}{2}}\r)^{2,0,0} \;.
\label{eq.su32}
\ee
Closure examination of the reduced matrix element $\lan \eta,\ell_f \mid\mid Q^2
\mid\mid \eta,\ell_i\ran$ of the quadrupole operator in the orbital space allows
us to recognize that there will be multiple $SU(3)$ subalgebras in $U(\can)$. We
will turn to this now.

As Elliott considered \cite{Ell-58a}, the quadrupole operator is $Q^2_q=
\sqrt{\frac{4\pi}{5}}\l[ r^2Y^2_q(\theta , \phi) + p^2Y^2_q(\theta_p ,
\phi_p)\r]$ with oscillator length parameter $b=1$. For a single shell. this is
equivalent to using $Q^2_q = \sqrt{\frac{16\pi}{5}}\,r^2Y^2_q(\theta ,\phi)$.
Therefore, the reduced matrix elements of $Q^2$ decompose into the radial part
and angular part,
\be
\lan \eta,\ell_f \mid\mid Q^2 \mid\mid \eta,\ell_i\ran = 
\lan \eta,\ell_f \mid\mid \dis\sqrt{\f{16\pi}{5}}Y^2(\theta,\phi) 
\mid\mid \eta, \ell_i
\ran\;\lan \eta,\ell_f \mid\mid r^2 \mid\mid \eta,\ell_i\ran\;,
\label{eq.su3a}
\ee 
with the angular part given by \cite{Brussard},
\be
\barr{l}
\lan \eta,\ell \mid\mid \dis\sqrt{\f{16\pi}{5}}Y^2(\theta,\phi) \mid\mid 
\eta, \ell\ran  = -2 \dis\sqrt{\dis\f{\ell(\ell+1)(2\ell+1)}{(2\ell+3)
(2\ell-1)}}\;,\\
\lan \eta,\ell \mid\mid \dis\sqrt{\f{16\pi}{5}}Y^2(\theta,\phi) \mid\mid 
\eta, \ell+2\ran  = \lan \eta,\ell+2 \mid\mid
\dis\sqrt{\f{16\pi}{5}}Y^2(\theta,\phi) \mid\mid \eta, \ell\ran 
= \dis\sqrt{\dis\f{6(\ell+1)(\ell+2)}{(2\ell+3)}}\;.
\earr \label{eq.su33}
\ee
Similarly, the radial matrix elements are
\be
\barr{l}
\lan \eta,\ell \mid\mid r^2 \mid\mid \eta,\ell\ran = \dis\frac{2\eta+3
}{2}\;,\\
\lan \eta,\ell \mid\mid r^2 \mid\mid \eta,\ell+2\ran = 
\lan \eta,\ell+2 \mid\mid r^2 \mid\mid \eta,\ell\ran = 
\alpha_{\ell,\ell+2}\dis\sqrt{(\eta-\ell)(\eta+\ell+3)}\;;\\
\alpha_{\ell,\ell+2} = \alpha_{\ell+2,\ell} = \pm 1\;.
\earr \label{eq.su34}
\ee
The phase factor $\alpha_{\ell,\ell+2}$ arises as there is freedom in choosing 
the phases of the radial wavefunctions of a 3D oscillator. In SM studies, the
standard convention is to use $\alpha_{\ell,\ell+2}=-1$ for all $\ell$
\cite{octup,Brussard,Bertsch}. However, Elliott in his $SU(3)$ introductory
paper \cite{Ell-58a} and in $sd$ as well as $sdg$ IBM and IBFM the choice made
is $\alpha_{\ell,\ell+2}=+1$ for all $\ell$ \cite{Iac-87,KY-rev,Piet,BK}. Thus,
in general we have,
\be
\barr{l}
L^1_q = 2 \dis\sum_\ell \dis\sqrt{\dis\frac{\ell(\ell+1)(2\ell+1)}{3}}
\l(a^\dagger_{\ell \f{1}{2} \f{1}{2}} \tilde{a}_{\ell 
\f{1}{2} \f{1}{2}}\r)^{1,0,0}_q \;,\\
Q^2_q(\baa)= -2(2\eta+3)\,\dis\sum_\ell
\dis\sqrt{\dis\frac{\ell(\ell+1)(2\ell+1)}{5(2\ell+3)(2\ell-1)}}
\l(a^\dagger_{\ell \f{1}{2} \f{1}{2}} \tilde{a}_{\ell \f{1}{2} 
\f{1}{2}}\r)^{2,0,0}_q \\
+ \dis\sum_{\ell <
\eta}\,2\alpha_{\ell,\ell+2}\;\dis\sqrt{\dis\frac{6(\ell+1)(\ell+2)(\eta-\ell)
(\eta+\ell+3)}{5(2\ell+3)}}\l[\l(a^\dagger_{\ell \f{1}{2} \f{1}{2}} 
\tilde{a}_{\ell+2, \f{1}{2} \f{1}{2}}\r)^{2,0,0}_q +
\l(a^\dagger_{\ell+2, \f{1}{2} \f{1}{2}} \tilde{a}_{\ell \f{1}{2} \f{1}{2}}
\r)^{2,0,0}_q \r]\;; \\
\baa=(\alpha_{0,2}, \alpha_{2,4}, \ldots, \alpha_{\eta-2,\eta})\;\;
\mbox{for}\;\;\eta\;\;\mbox{even}\;,\\
\baa=(\alpha_{1,3}, \alpha_{3,5}, \ldots, \alpha_{\eta-2,\eta})\;\;
\mbox{for}\;\;\eta\;\;\mbox{odd}\;,\\
\baa=(\pm 1, \pm 1, \ldots)\;.
\earr \label{eq.su35}
\ee  
Now, the most important result that can be proved by using the tedious but
straight forward angular momentum algebra is that the eight operators $(L^1_q ,
Q^2_{q^\prime}(\baa))$ generate $SU(3)$ algebra independent of the choice of the
$\baa$'s and they satisfy the commutation relations \cite{Ell-58a,octup},
\be
\barr{l}
\l[L^1_q\,,\,L^1_{q^\pr}\r] = -\dis\sqrt{2} \lan 1 q\;1 q^\pr \mid 1 q+q^\pr
\ran\,L^1_{q+q^\pr}\;,\\
\l[L^1_q\,,\,Q^2_{q^\pr}(\baa)\r] = -\dis\sqrt{6} \lan 1 q\;2 q^\pr 
\mid 2 q+q^\pr \ran\,Q^2_{q+q^\pr}(\baa)\;,\\
\l[Q^2_q(\baa)\,,\,Q^2_{q^\pr}(\baa)\r] = 3\dis\sqrt{10} \lan 2 q\;2 q^\pr 
\mid 1 q+q^\pr\ran\,L^1_{q+q^\pr} \;.
\earr \label{eq.su36}
\ee
Thus, we have multiple $SU(3)$ algebras $SU^{\baa}(3)$ in SM spaces generated by
the operators in Eq. (\ref{eq.su35}). Clearly for a given $\eta$, there will be
$2^{\l[\f{\eta}{2}\r]}$ number of $SU(3)$ algebras; $\l[\f{\eta}{2}\r]$ is the
integer part of $\eta/2$.  Then, we have two $SU(3)$ algebras in $sd$ ($\eta=2$)
and $pf$ ($\eta=3$) shells, four $SU(3)$ algebras in $sdg$ ($\eta=4$) and $pfh$
($\eta=5$) shells, eight $SU(3)$ algebras in $(sdgi)$ ($\eta=6$) and $(pfhj)$
($\eta=7$) shells and so on. Thus, the first non-trivial situation that is not
discussed in literature before is $sdg$ or $\eta=4$ shell with four $SU(3)$
algebras $SU^{(-,-)}(3)$, $SU^{(+,-)}(3)$, $SU^{(-,+)}(3)$ and $SU^{(+,+)}(3)$.
Here, $\baa=(\alpha_{sd},\alpha_{dg})$ and $(-,-)$ means
$(\alpha_{sd},\alpha_{dg})=(-1,-1)$ and similarly for other choices of
$(\alpha_{sd},\alpha_{dg})$. In the reminder of this paper, we will use the
example of $\eta=4$ shell to present some results from multiple $SU(3)$
algebras. Before this, we will first consider the quadrupole-quadrupole
interaction generated by $Q^2_q(\baa)$. 

\subsection{Matrix elements of Quadrupole-quadrupole interaction from multiple
$SU(3)$ algebras} 

Investigation of multiple $SU(3)$ algebras in shell model spaces needs firstly
the single particle energies (spe) and two-body matrix elements (TBME) of the
quadrupole-quadrupole interaction operator $Q^2(\baa) \cdot Q^2(\baa)$ for all
phase choices $\baa$ (also the spe and TBME for the simpler $L \cdot L$
operator). The methods for obtaining these are well known \cite{Brussard} and we
will give only the final formulas. In order to derive formulas for the spe and
TBME generated by $Q^2(\baa) \cdot Q^2(\baa)$ operators, firstly notice that the
$Q^2_q$  operator can be written as,
\be
Q^2_q(\baa) = 2\dis\sum_{\ell_f , \ell_i} C^{\baa}_{\ell_f , \ell_i} 
\l(a^\dagger_{\ell_f \f{1}{2} \f{1}{2}}
\tilde{a}_{\ell_i \f{1}{2} \f{1}{2}}\r)^{2,0,0}_q\;.
\label{eq.qq1}
\ee
The $C^{\baa}_{\ell_f , \ell_i}$ follow easily from Eq. (\ref{eq.su35}). From 
now on we will drop '2' and $\baa$ in $Q^2_q(\baa)$ when there is no confusion. 
For a many particle system, 
\be
Q \cdot Q = \sum_{i=1}^m Q(i) \cdot Q(i) + 2 \sum_{i<k=1}^m Q(i) \cdot 
Q(k)
\label{eq.qmany}
\ee
where $i$ and $k$ are particle indices and $m$ is number of particles.  The
first sum generates spe and the second term TBME. Given the shell model single
particle $(n \ell j)$-orbits (note that the oscillator shell number
$\eta=2n+\ell$), matrix elements of $Q(1) \cdot Q(2)$ in the two-particle 
antisymmetric states (called a.s.m.)  can be written in terms of the matrix
elements in the two-particle non-antisymmetric states (called n.a.s.m.) as,
\be
\barr{l}
\lan (j_a j_b)JT \mid Q(1) \cdot Q(2) \mid(j_c j_d)JT\ran_{a.s.m.} = \\
\dis\frac{\lan (j_a j_b)JT \mid Q(1) \cdot Q(2) \mid(j_c j_d)JT\ran_{n.a.s.m.} +
(-1)^{J+T-j_c -j_d} \lan (j_a j_b)JT \mid Q(1) \cdot Q(2) \mid(j_d j_c)JT\ran_{
n.a.s.m.}}{\dis\sqrt{\l(1+\delta_{ab}\r) \l(1+\delta_{cd}\r)}}\;.
\earr\label{eq.qq2}
\ee
Using angular momentum algebra it is easy to recognize that,
\be
\barr{l}
\lan (j_a j_b)JT \mid Q(1) \cdot Q(2) \mid(j_c j_d)JT\ran_{n.a.s.m.} = 
(-1)^{j_b + j_c +J} \l\{\barr{ccc} j_a & j_b & J\\ j_d & j_c & 2\earr\r\} \\
\times\;\lan j_a \mid\mid Q \mid\mid j_c\ran\,\lan j_b \mid\mid Q \mid\mid 
j_d\ran\;.
\earr \label{eq.qq3}
\ee
The reduced matrix elements $\lan \mid\mid Q \mid\mid \ran$ are given by, 
\be
\barr{l}
\lan \eta, \ell_f, j_f \mid\mid Q^2(\baa) \mid\mid \eta, \ell_i, j_i\ran = 
(-1)^{\ell_f + \frac{1}{2} + j_i + 2} \\
\times\;\dis\sqrt{5 (2j_i +1)(2j_f +1)}\;\l\{\barr{ccc}\ell_f & j_f &
\frac{1}{2} \\ j_i & \ell_i & 2 \earr\r\}\;C^{\baa}_{\ell_f , \ell_i} \;.
\earr \label{eq.qq4}
\ee
Combining Eqs. (\ref{eq.qq3}) and (\ref{eq.qq4}) with Eq. (\ref{eq.qq2}) and 
Eq. (\ref{eq.qmany}) will give the TBME of the $Q^2(\baa) \cdot Q^2(\baa)$ 
operator. The spe $\epsilon^{\baa}_{\ell j}$ of the $Q^2(\baa) \cdot Q^2(\baa)$
are simply given by
\be
\epsilon^{\baa}_{\ell j} = \dis\frac{5}{2\ell +1}\; \dis\sum_{\ell^\prime} 
\l|C^{\baa}_{\ell \ell^\prime}\r|^2\;.
\ee
An important property of the $Q^2(\baa) \cdot Q^2(\baa)$ operator is that it is
related to the quadratic Casimir invariant ($C_2$) of $SU^{\baa}(3)$ in a
simple manner,
\be
-Q^2(\baa) \cdot Q^2(\baa) = -C_2(SU^{\baa}(3)) + \dis\f{3}{4} L \cdot L\;.
\label{eq.su37}
\ee
The procedure described above will also give the spe and TBME of $L \cdot L$
operator. Let us mention that the eigenvalue of $C_2(SU^{\baa}(3))$ over a
$SU^{\baa}(3)$ irrep $(\la \mu)$ is $\la^2 + \mu^2 + \la \mu +3(\la +\mu)$.
Also, note that the dot product in Eqs. (\ref{eq.su37}) and (\ref{eq.qmany}) is
with respect to the  orbital space.

\subsection{Correlation between different $Q \cdot Q$ operators}

In order to gain some insight into the differences between different
$SU^{\baa}(3)$ algebras, we will consider the correlation in $m$ nucleon spaces
between different $Q(\baa) \cdot Q(\baa)$ operators. For this, we will use the
example of $\eta=4$ shell giving $(n \ell j)$ to be $(2,0,1/2)$, $(1,2,3/2)$,
$(1,2,5/2)$, $(0,4,7/2)$ and $(0,4,9/2)$. In this space, spe and TBME are
obtained for $Q^2(\baa) \cdot Q^2(\baa)$ operators with
$\baa=(\alpha_{sd},\alpha_{dg}) = (+,+), (+,-), (-,+)$ and $(-,-)$ using the
results in Section IIA.

Given an operator $\co$ acting in $m$ particle spaces ($\co$ is assumed to be 
real), its trace over the $m$ particle space is $\lan\lan \co \ran\ran^m =
\sum_\gamma\,\lan m, \gamma \mid \co \mid m,\gamma\ran$. Note that $\l.\l|
m,\gamma\r.\ran$ are $m$-particle states. Similarly, the $m$-particle average is
$\lan \co\ran^m =[d(m)]^{-1} \lan\lan \co \ran\ran^m$ where $d(m)$ is
$m$-particle space dimension. Using the spectral distribution method of French
\cite{CFT,KH-10}, a geometry can be defined \cite{CFT} with norm (or size or
length) of an operator $\co$ given by $\mid\mid \co \mid\mid_m = \sqrt{\lan
\tilde{\co}\tilde{\co} \ran^m}$; $\tilde{\co}$ is the traceless part of $\co$.
Following this, given any two operators $\co_1$ and $\co_2$, the correlation
coefficient 
\be
\zeta(\co_1,\co_2)= \dis\frac{\lan \widetilde{\co_1} \widetilde{\co_2} \ran^m}{
\mid\mid \co_1 \mid\mid_m \;\mid\mid \co_2 \mid\mid_m}\;,
\label{eqss1}
\ee
gives the cosine of the angle between the two operators. Thus, $\co_1$ and
$\co_2$ are same within a normalization constant if $\zeta=1$ and they are
orthogonal to each other if $\zeta=0$ \cite{KH-10}. Most recent
application of norms and correlation coefficients is in understanding
the structure of multiple pairing algebras in shell model \cite{Ko-BJP}.

Applying Eq. (\ref{eqss1}), we have calculated $\zeta$ between the operators
$Q^2(\alpha_{sd},\alpha_{dg}) \cdot Q^2(\alpha_{sd},\alpha_{dg})$ and
$Q^2(\alpha^\prime_{sd},\alpha^\prime_{dg}) \cdot
Q^2(\alpha^\prime_{sd},\alpha^\prime_{dg})$ for all possible combinations of
$\alpha$'s and $(\alpha^\prime)$'s. Some results for $\zeta$ are given in Table
I. It is seen from the table that $Q^2(-,-) \cdot Q^2(-,-)$ is strongly
correlated with $Q^2(+,-) \cdot Q^2(+,-)$. Similarly, the $Q \cdot Q$'s with 
$(\alpha_{sd},\alpha_{dg})=(+,+)$ and $(-,+)$ are strongly correlated. However,
the correlations between other pairs of $Q^2 \cdot Q^2$ are quite small. Thus,
$SU^{(-,-)}(3)$ and $SU^{(+,-)}(3)$ are expected to give similar results but
quite different from  $SU^{(+,+)}(3)$ and $SU^{(-,+)}(3)$. This is seen in the
results of detailed calculations presented in the next section. It is important
to stress that all the four $SU^{\baa}(3)$ algebras generate the same spectrum 
for $H(\baa) = Q^2(\alpha_{sd},\alpha_{dg}) \cdot Q^2(\alpha_{sd},\alpha_{dg})$
independent of $(\alpha_{sd},\alpha_{dg})$.  We will consider these in more
detail in the following.

\begin{table*}

\caption{Correlation coefficient $\zeta$ between $Q \cdot Q$ operators with
different values for the phases $(\alpha_{sd}, \alpha_{dg})$ in $sdg$ shell
model $m$-particle spaces ($m$ is number of nucleons). Note that the total
number of single particle states (with spin and isospin) is 60. The $\zeta$
values in column 3 are for $m=4$, 8, 12, 20, 30, 40, 50 and 56. See text for
other details.}

\begin{tabular}{cc|l}
\hline 
\hline
$(\alpha_{sd},\alpha_{dg})$ & $(\alpha^\prime_{sd},\alpha^\prime_{dg})$  & 
$\zeta$ \\
\hline
\hline
$(-,-)$ & $(+,-)$ & $0.39,0.36,0.35,0.35,0.35,0.35,0.36,0.39$\\
& $(-,+)$ & $0.14,0.1,0.08,0.08,0.08,0.08,0.09,0.14$ \\
& $(+,+)$ & $0.07,0.02,0.01,0.0,0.0,0.0,0.01,0.07$ \\
$(+,-)$ & $(-,+)$ & $0.07,0.02,0.01,0.0,0.0,0.0,0.01,0.07$ \\
& $(+,+)$ & $0.14,0.1,0.08,0.08,0.08,0.08,0.09,0.14$ \\
$(-,+)$ & $(+,+)$ & $0.39,0.36,0.35,0.35,0.35,0.35,0.36,0.39$ \\
\hline\hline
\end{tabular}
\label{corr}
\end{table*}

\section{Results for Spectra, quadrupole moments and $E2$ transition strengths 
from SM and DSM}

With the $sdg$ example, we have four $Q \cdot Q$ Hamiltonians,
\be
\barr{rcl}
H^{(-,-)}_Q = - Q^2(-,-) \cdot Q^2(-,-)\;,\\
H^{(+,-)}_Q = - Q^2(+,-) \cdot Q^2(+,-)\;,\\
H^{(-,+)}_Q = - Q^2(-,+) \cdot Q^2(-,+)\;,\\
H^{(+,+)}_Q = - Q^2(+,+) \cdot Q^2(+,+)\;.
\earr \label{eq.su38}
\ee
In this section we will present the results generated by these four $H$'s for
the yrast levels, quadrupole moments $Q_2(J)$ of these levels and the $B(E2)$'s
along the yrast line for $J$ up to 10. Used for this purpose are the Antoine
shell model code \cite{Anton} and also the deformed shell model (DSM) based on
Hartree-Fock states \cite{KS}. DSM is particularly important for bringing out
shape information in a transparent manner and also it is useful for larger
particle numbers where SM calculations are impractical. We will test the SM
results with analytical results derived using $SU(3)$ algebra and also test DSM
using SM results. We will first present some analytical results from $SU(3)$
algebra.

\begin{table}
\begin{center}

\caption{Ground state or leading $SU(3)$ irrep $(\lambda_H , \mu_H)$ for a
given  number $m$ of identical nucleons and also for a given number $m$ of 
nucleons with isospin $T=|T_Z|$. Results are given for the oscillator  shell
$\eta=4$. The   $(\lambda_H , \mu_H)$ are given in the table as $(\lambda_H ,
\mu_H)^m$ for identical nucleons with $m \geq 2$ and $(\lambda_H ,\mu_H)^{m,T}$
for nucleons with $T=|T_z|$ and $3 \leq m \leq 15$; for odd $m$ values, $2T$
value given instead of $T$ value. More complete results are available in
\cite{Kota-hw}.}
\begin{tabular}{l}
\hline
$\eta=4$: identical  nucleons \\
$( 8, 0)^ 2$,$(10, 1)^ 3$,$(12, 2)^ 4$,$(15, 1)^ 5$,$(18, 0)^ 6$,$(18, 2)^ 7$,
$(18, 4)^ 8$,$(19, 4)^ 9$,$(20, 4)^{10}$,$(22, 2)^{11}$,$(24, 0)^{12}$,\\
$(22, 3)^{13}$,
$(20, 6)^{14}$,$(19, 7)^{15}$,$(18, 8)^{16}$,$(18, 7)^{17}$,$(18, 6)^{18}$,
$(19, 3)^{19}$,$(20, 0)^{20}$,$(16, 4)^{21}$,$(12, 8)^{22}$,\\
$( 9,10)^{23}$,$( 6,12)^{24}$,
$( 4,12)^{25}$,$( 2,12)^{26}$,$( 1,10)^{27}$,$( 0, 8)^{28}$,$( 0, 4)^{29}$,
$( 0, 0)^{30}$\\
$\eta=4$: even number of nucleons  \\
$(16, 0)^{ 4, 0}$,$(14, 1)^{ 4, 1}$,$(12, 2)^{ 4, 2}$,$(20, 2)^{ 6, 0}$,
$(20, 2)^{ 6, 1}$,$(19, 1)^{ 6, 2}$,
$(18, 0)^{ 6, 3}$,$(24, 4)^{ 8, 0}$,$(25, 2)^{ 8, 1}$,\\
$(26, 0)^{ 8, 2}$,$(22, 2)^{ 8, 3}$,$(18, 4)^{ 8, 4}$,
$(30, 2)^{10, 0}$,$(30, 2)^{10, 1}$,$(28, 3)^{10, 2}$,$(26, 4)^{10, 3}$,
$(23, 4)^{10, 4}$,$(20, 4)^{10, 5}$,\\
$(36, 0)^{12, 0}$,$(33, 3)^{12, 1}$,$(30, 6)^{12, 2}$,$(29, 5)^{12, 3}$,
$(28, 4)^{12, 4}$,$(26, 2)^{12, 5}$,
$(24, 0)^{12, 6}$,$(36, 4)^{14, 0}$,$(36, 4)^{14, 1}$,\\
$(34, 5)^{14, 2}$,$(32, 6)^{14, 3}$,$(32, 3)^{14, 4}$,
$(32, 0)^{14, 5}$,$(26, 3)^{14, 6}$,$(20, 6)^{14, 7}$ \\
$\eta=4$: odd number of nucleons \\
$(12, 0)^{ 3, 1}$, $(10, 1)^{ 3, 3}$, $(18, 1)^{ 5, 1}$, $(16, 2)^{ 5, 3}$, 
$(15, 1)^{ 5, 5}$, $(22, 3)^{ 7, 1}$, $(23, 1)^{ 7, 3}$, $(22, 0)^{ 7, 5}$, \\
$(18, 2)^{ 7, 7}$, $(27, 3)^{ 9, 1}$, $(28, 1)^{ 9, 3}$, $(26, 2)^{ 9, 5}$, 
$(22, 4)^{ 9, 7}$, $(19, 4)^{ 9, 9}$, $(33, 1)^{11, 1}$, $(30, 4)^{11, 3}$, \\
$(28, 5)^{11, 5}$, $(27, 4)^{11, 7}$, $(24, 4)^{11, 9}$, $(22, 2)^{11,11}$, 
$(36, 2)^{13, 1}$, $(33, 5)^{13, 3}$, $(31, 6)^{13, 5}$, $(30, 5)^{13, 7}$, \\
$(30, 2)^{13, 9}$, $(28, 0)^{13,11}$, $(22, 3)^{13,13}$, $(36, 6)^{15, 1}$, 
$(37, 4)^{15, 3}$, $(35, 5)^{15, 5}$, $(34, 4)^{15, 7}$, $(34, 1)^{15, 9}$, \\
$(30, 3)^{15,11}$, $(24, 6)^{15,13}$, $(19, 7)^{15,15}$ \\
\hline
\end{tabular}
\end{center}
\end{table}

\subsection{Analytical results from $SU(3)$ algebra}

With $SU(3)$ symmetry of the $H_Q$ Hamiltonians, the shell model space for a $m$
nucleon system decomposes into $SU(3)$  irreducible representations (irreps) due
to the equivalence between $H_Q$ and $C_2(SU(3))$ as given by  Eq.
(\ref{eq.su37}). If we have identical nucleons (protons or neutrons), the ground
band belongs to the leading $SU(3)$ irrep $(\la_H , \mu_H)$ with spin $S=0$ and
$J=L$ for even $m$ (similarly with $S=1/2$ for odd $m$). It is easy to write a
formula for obtaining $(\la_H , \mu_H)$ as given in \cite{Kota-hw}. The irreps
for $m$ identical nucleons in $\eta=4$ shell are given in Table II. Similarly,
for $m$ nucleons with isospin $T$, we need to consider the lowest spin-isospin
$SU(4)$ irrep allowed for this system \cite{Manan,Piet-su42}  and this will then
give $(\la_H , \mu_H)$ \cite{Kota-hw}. The irreps $(\la_H , \mu_H)$ for $m$
nucleons with $T=|T_z|$ are given in Table II. The eigenstates of $H_Q$ are
$\l|m;(\la_H \mu_H)KL;S:JT\ran$ and the $(\la_H \mu_H) \rightarrow L$ reduction
is well known giving,
\be
\barr{rcl}
(\lambda \mu) \longrightarrow L: & & \\
K & = & min(\lambda,\mu),\; min(\lambda,\mu) -2,\; \cdots , 0\;\mbox{or}\;1, \\
L &=& K,\; K+1, \;K+2,\; \cdots , \; K + max(\lambda,\mu)\;\; for\;\;
K \neq 0\;, \\
L & = & max(\lambda,\mu),\;max(\lambda,\mu)-2,\; \cdots , 0\;\mbox{or}\; 
1\;\;for\;\; K = 0\;,
\\
(\lambda,\mu) \rightarrow L &\Longleftrightarrow& (\mu,\lambda) 
\rightarrow L \;.
\earr \label{eq.su39}
\ee
It is easy to see that the energies of the yrast levels in a even $m$ system
(assuming spin $S=0$) are given by, 
\be
E(J=L) = -(\la_H^2 +\mu_H^2 +\la_H\mu_H +3(\la_H +\mu_H)) + \f{3}{4}L(L+1)\;.
\label{eq.su310}
\ee
In the examples presented ahead in the present paper we will only consider even
$m$ systems with $(\la_H \mu_H)=(\la 0)$ and then $\la$ is even. A $(\la ,0)$
irrep with $\la$ even, as seen from Eq. (\ref{eq.su39}), generates the ground
band with $J=0$, $2$, $4$, $\ldots$, $\lambda$. The ground state energy
$E_{gs}=(\la^2 +3\la)$ and the energies of the $J$ levels with respect to
$E_{gs}$ are just $3J(J+1)/4$. In addition, if we choose the $E2$ transition
operator to be the $Q$ of one of the $H_Q$, then formulas for $Q_2(J)$ and
$B(E2)$ will be simple for the $(\la, 0)$ irrep of the corresponding $SU(3)$
algebra. Just as it is considered in SM and DSM codes, we will take the $E2$
operator $T^{E2}$ for identical nucleon systems to be 
\be
T^{E2} = Q^2_q(-,-)\; e_{eff}\, b^2
\label{eq.su311} 
\ee
where $b$ is the oscillator length parameter and $e_{eff}$ is effective charge.
Then, analytical formulas for the quadrupole moments $(Q(J))$ of the yrast
levels and  $B(E2)$'s among them follow  from the simple $SU(3)$ algebra for the
eigenstates obtained for $H_Q^{(-,-)}$ as they belong to $SU^{(-,-)}(3)$. Using
the results in \cite{Ell-58a,BK},  we have for $H_Q^{(-,-)}$ in Eq.
(\ref{eq.su38}) with $T^{E2}$ in Eq.  (\ref{eq.su311}),
\be
\barr{l}
Q((\la, 0):J=L) = -\dis\f{L}{2L+3} (2\la +3)\,e_{eff} b^2\;,\\ 
B(E2;(\la ,0) J=L \rightarrow J-2=L-2) = \dis\frac{5}{16\pi} \l\{
\dis\frac{6J(J-1)(\la -J+2)(\la +J+1)}{(2J-1)(2J+1)}\r\}(e_{eff})^2 b^4\;.
\earr \label{eq.su312}
\ee
However, for systems with valence protons and neutrons, the $E2$ transition
operator is taken to be
\be
T^{E2} = \l[e^p_{eff}\,Q^2_q(-,-;p) + e^n_{eff}\,Q^2_q(-,-;n)\r]\,  b^2
\label{eq.e2qpqn}
\ee 
where $e^p_{eff}$ and $e^n_{eff}$ are proton and neutron effective charges.
Again, using eigenstates obtained for $H_Q^{(-,-)}$ as they belong to
$SU^{(-,-)}(3)$ and the $T^{E2}$ in Eq. (\ref{eq.e2qpqn}), a simple formula 
is obtained for $Q(J)$ and $B(E2)$'s in the
situation where the ground band is given by $\l|(\la_\pi , 0)(\la_\nu ,
0)(\la_\pi + \la_\nu , 0)K=0,L,S=0,J=L\ran$ for a system with protons ($\pi$) and
neutrons ($\nu$). Now, carrying out the $SU(3)$ algebra using the mathematical
formulation and analytical results given in \cite{JPD-1,Hecht-65,Mill,Verg} we
have,  
\be
\barr{l}
Q((\la , 0):J=L) = -\dis\f{L}{2L+3} 2(\la +3)\,X_{eff}\, b^2\;,\\ 
B(E2;(\la ,0) J=L \rightarrow J-2=L-2) = \dis\frac{5}{16\pi} \l\{
\dis\frac{6J(J-1)(\la -J+2)(\la +J+1)}{(2J-1)(2J+1)}\r\}(X_{eff})^2\, b^4\;; \\
X_{eff}=\dis\frac{e^p_{eff}\l(\la_\pi^2 + 3\la_\pi +\la_\pi \la_\nu
\r) + e^n_{eff}\l(\la_\nu^2 + 3\la_\nu +\la_\pi \la_\nu\r)}{\l(\la^2 +3\la\r)}\;,
\;\; \la=\la_\pi + \la_\nu \;.
\earr \label{eq.e2formu}
\ee
Tests of Eqs. (\ref{eq.su310}), (\ref{eq.su312}) and (\ref{eq.e2formu}) are 
carried out using SM and DSM in the next three subsections. 

It is important to stress that in the event we use the eigenstates of other
$H^{\baa}_Q$, the ground band generated by them will belong to the $(\la 0)$
irrep of the corresponding $SU^{\baa}(3)$. However, then the $Q$'s in $T^{E2}$
in Eqs. (\ref{eq.su311}) and (\ref{eq.e2qpqn}) are no longer generators of these
$SU^{\baa}(3)$'s and hence the formulas in Eqs. (\ref{eq.su312}) and
(\ref{eq.e2formu}) will not apply. In this situation, we have to use $Q^2_q(-,-)
= Q^2_q(\baa) +\Delta Q$ and $\Delta Q$ follows easily from Eq. (\ref{eq.su35}).
Then, one has to carry out the $SU(3)$ tensorial decomposition of $\Delta Q$
with respect to $SU^{\baa}(3)$ and use the $SU(3)$ Wigner-Racah algebra as
described for example in \cite{JPD-1,Hecht-65,Mill,Verg} for obtaining the
matrix elements of $\Delta Q$ in the $\l|(\la 0)K=0,L\ran$ states. This exercise
is postponed to a future publication and instead we will present results of full
(without any truncation) SM results along with some DSM results in the next two
subsections and only DSM results in the third subsection. In addition, to gain
more insight into the other $SU^{\baa}(3)$ algebras, we  will use the asymptotic
formulas for quadrupole moments and $B(E2)$'s in $sdg$IBM in Section IV.

\begin{table}
\begin{center}
\caption{Shell model results for quadrupole moments 
$Q(J)$ and $B(E2; J \rightarrow
J-2)$ values for the ground $K=0^+$ band members for a system of 6 protons in 
$\eta=4$ shell. Results are given for the four Hamiltonians in 
Eq. (\ref{eq.su38}). In the table $(-,-)$ means we are using the wavefunctions 
obtained using $H_Q^{(-,-)}$ and similarly others. For other details see text.}
\begin{tabular}{ccccc}
\hline
$J$ & \multicolumn{4}{c}{$Q(J)\; efm^2$} \\
& $(-,-)$ & $(+,-)$ & $(-,+)$ & $(+,+)$ \\
\hline
$2^+_1$ & $-49.18$ & $-33.90$ & $-1.85$ & $13.44$ \\
$4^+_1$ & $-62.59$ & $-40.16$ & $-4.71$ & $17.72$ \\
$6^+_1$ & $-68.85$ & $-39.78$ & $-9.12$ & $19.97$ \\
$8^+_1$ & $-72.48$ & $-37.06$ & $-14.96$ & $20.46$ \\
$10^+_1$ & $-74.84$ & $-34.57$ & $-21.99$ & $18.28$\\
\hline
$J$ & \multicolumn{4}{c}{$B(E2; J \rightarrow J-2)\; e^2fm^4$} \\
& $(-,-)$ & $(+,-)$ & $(-,+)$ & $(+,+)$ \\
\hline
$2^+_1$ & $585.97$ & $291.34$ & $0.42$ & $42.20$ \\ 
$4^+_1$ & $815.31$ & $388.79$ & $1.38$ & $58.58$ \\
$6^+_1$ & $853.68$ & $377.33$ & $3.90$ & $61.14$ \\
$8^+_1$ & $827.24$ & $325.68$ & $9.24$ & $58.93$\\
$10^+_1$ & $760.52$ & $254.56$ & $18.38$ & $53.81$ \\
\hline
\end{tabular}
\end{center}
\end{table}   
\begin{figure}[ht]
\begin{center}
\includegraphics[width=0.5\linewidth]{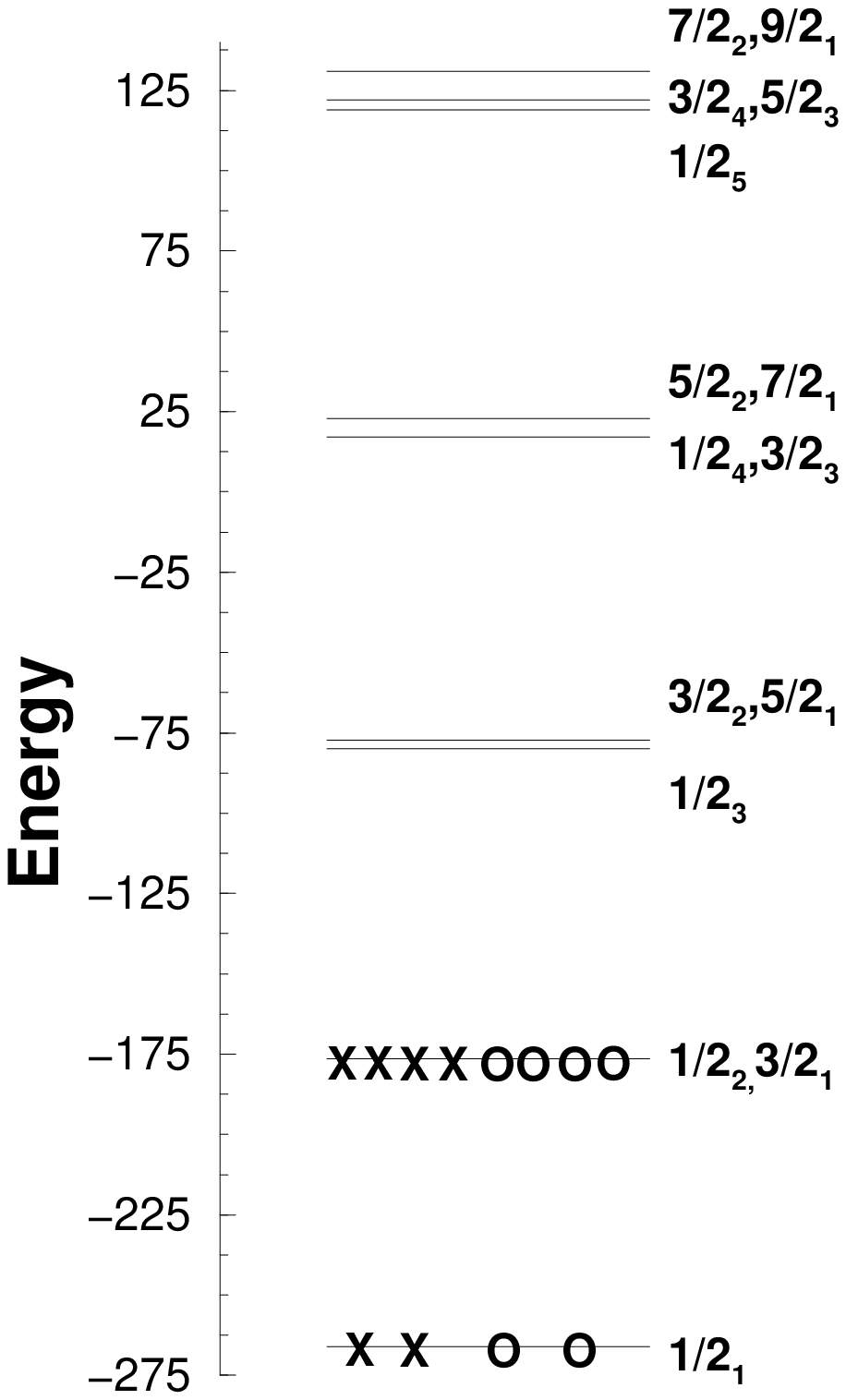}
\end{center}

\caption{Hartree-Fock sp spectrum (it is same for both protons and neutrons) and
the lowest intrinsic state for the  $(sdg)^{6_p , 6_n}$ system generated by the
four $H_Q$ operators in Eq. (\ref{eq.su38}).  In the figure, the symbol $\times$
denotes neutrons and $0$ denotes protons. Shown in the figure are the $k$ values
of the sp orbits and each orbit is doubly degenerate with $\l|k\ran$ and
$\l|-k\ran$ states. The spectrum is same for all  the four Hamiltonians although
the sp wavefunctions are different. The HF energy $E_{HF}$ for the lowest
intrinsic state is -1351.73 (note that $E$ is unit less and the unit MeV has to
be put back after multiplying with an appropriate scale factor if the results
are used for a real nucleus) for all the four Hamiltonians.  The intrinsic
quadrupole moments  (in units of $b^2$), calculated using
$T^{E2}=Q^2_q(-,-)\,b^2$ as the  quadrupole operator,   for $H_Q^{(-,-)}$,
$H_Q^{(+,-)}$, $H_Q^{(-,+)}$ and $H_Q^{(+,+)}$ are $71.95$, $47.43$, $5.06$ and
$-19.45$ respectively. See text for other details. }

\label{fig1}
\end{figure}

\subsection{SM and DSM results for multiple $SU(3)$ algebras: $(sdg)^{6_p}$
example}

In our first example, we have analyzed a system of 6 protons in $\eta=4$ shell,
i.e. $(sdg)^{6_p}$ system by carrying out SM calculations  using the four $H_Q$
Hamiltonians in the full SM space (matrix dimension in the $m$-scheme is $\sim
10^5$) using the Antoine code. For this system, the leading $SU(3)$ irrep (see
Table II) is $(18,0)$ with $S=0$. Then, Eq. (\ref{eq.su310}) gives $E_{gs}=378$
and SM calculations for all four $H_Q$'s are in agreement with this $SU(3)$
result. Also, in the SM results the excitation energies of the yrast $J$  states
or ground band members ($J=0,,2,4,6,\ldots$) are seen to follow for all  the
four $H_Q$'s the $3J(J+1)/4$ law as given by $SU(3)$.  Thus, it is verified by
explicit SM calculations that all the four $H_Q$'s give $SU(3)$ symmetry. Though
the energy spectra are same, the wavefunctions of the yrast $J$ states are
different. This is established by calculating $Q(J)$ and $B(E2)$'s for the
ground band members using $T^{E2}$ given by Eq. (\ref{eq.su311}). In all the
calculations, $e_{eff}=1e$ and $b^2=A^{1/3}\,fm^2$ with $A=86$ are used. 
Results from SM for the four $H_Q$'s are given in Tables III. It is easy to see
that the results for $H_Q^{(-,-)}$ are in complete agreement with the results
the $SU(3)$ formulas given by Eq. (\ref{eq.su312}). This is expected as $T^{E2}$
in Eq. (\ref{eq.su311}) is a generator of $SU^{(-,-)}(3)$ generated by
$H_Q^{(-,-)}$. However, the results from the other three $H_Q$'s are quite
different and do not follow the $SU(3)$ results in Eq. (\ref{eq.su312}) as the
$T^{E2}$ chosen is not a generator of the $SU(3)$'s generated by the three
$H_Q$'s. It is seen from Tables III that the results for $Q(J)$ and $B(E2)$'s
from $H_Q^{(+,-)}$ are closer to those from $H_Q^{(-,-)}$ and this is consistent
with the correlation coefficients shown in Table II. The $B(E2)$'s from
$H_Q^{(-,+)}$ are much smaller in magnitude. Moreover, $H_Q^{(-,-)}$ generates
prolate shape  and $H_Q^{(+,+)}$ oblate as seen clearly from Table III.
Quadrupole moments show that $H_Q(+,-)$ and $H_Q(-,+)$ also generate prolate
shapes but the deformation from $H_Q(-,+)$ is quite small for the low-lying
levels. To gain more insight into these results, we have performed DSM
calculations using the four $H_Q$'s with results as follows.

Starting with the same model space, sp energies and two-body interaction, in DSM
one solves Hartree-Fock (HF) sp  equations self-consistently assuming axial
symmetry. The lowest-energy prolate or oblate intrinsic state  for the nucleus
in question is then obtained. The various excited intrinsic states then are
obtained by making particle-hole ($p$-$h$) excitations over the lowest-energy
intrinsic state (lowest configuration). Carrying out angular momentum projection
from each intrinsic state and performing band mixing, orthonormalized
$\l|JK\ran$ states are obtained. See \cite{KS} for full details and many
applications of DSM. Latest application of DSM is to dark matter studies
\cite{Sahu}. In the present DSM calculations, only the lowest intrinsic state is
considered. It is found that the four $H_Q$'s generate the same  HF sp spectrum
and it is same as shown in Fig. 1 ahead except for some scale factors. The
lowest intrinsic state is obtained by putting two protons each in the $1/2_1$,
$1/2_2$ and $3/2_1$ states.  The intrinsic quadrupole moments (in units of
$b^2$) for $H_Q^{(-,-)}$, $H_Q^{(+,-)}$, $H_Q^{(-,+)}$ and $H_Q^{(+,+)}$ are
$+35.85$, $24.4$, $1.83$ and $-9.63$ respectively. Thus,   $H_Q^{(-,-)}$
generates prolate shape and $H_Q^{(+,+)}$ generates  oblate shape in agreement
with SM. It is important to emphasize that the intrinsic quadrupole moments are
calculated using $T^{E2}=Q^2_q(-,-)\,b^2$ as the  quadrupole operator.  The
ground state energy for the 6 proton system is found to be, for all the four
$H_Q$'s same as the exact $SU(3)$ values within less than 1\% deviation. The
energies of the yrast $J$ states from the ground state are also same for four
$H_Q$'s and they follow the $3J(J+1)/4$ law. Similarly, the results for $Q(J)$'s
and $B(E2)$'s are essentially same as the SM values. For example for
$H_Q^{(-,-)}$, the $Q(J)$ values (in $efm^2$ unit) are $-49.04$, $-62.38$,
$-68.56$, $-72.07$ and $-74.29$ for $J=2$, $4$, $6$, $8$ and $10$ respectively.
The corresponding $B(E2; J \rightarrow J-2)$ values (in $e^2fm^4$ unit) are
$582.78$, $810.44$, $848.78$, $822.26$ and $755.63$ respectively. Thus, for
larger particle systems where SM calculations are not possible, one can use with
confidence DSM for further insight into the results from the four $H_Q$'s, i.e.
from multiple $SU(3)$ algebras and this is used in Section III-D.

\begin{table}
\begin{center}
\caption{Shell model results for quadrupole moments 
$Q(J)$ and $B(E2; J \rightarrow
J-2)$ values for the ground $K=0^+$ band members for a 
system of 6 protons and 2 neutrons in 
$\eta=4$ shell. Results are given for the four Hamiltonians in 
Eq. (\ref{eq.su38}). In the table $(-,-)$ means we are using the wavefunctions 
obtained using $H_Q^{(-,-)}$ and similarly others. For other details see text.}
\begin{tabular}{ccccc}
\hline
$J$ & \multicolumn{4}{c}{$Q(J)\; efm^2$} \\
& $(-,-)$ & $(+,-)$ & $(-,+)$ & $(+,+)$ \\
\hline
$2^+_1$ & $-83.34$ & $-50.54$ & $-8.84$ & $23.96$ \\
$4^+_1$ & $-106.07$ & $-62.96$ & $-12.96$ & $30.15$ \\
$6^+_1$ & $-116.68$ & $-67.10$ & $-17.17$ & $32.42$ \\
$8^+_1$ & $-122.82$ & $-67.95$ & $-22.17$ & $32.70$ \\
$10^+_1$ & $-126.82$ & $-67.38$ & $-28.13$ & $32.32$ \\
\hline
$J$ & \multicolumn{4}{c}{$B(E2; J \rightarrow J-2)\; e^2fm^4$} \\
& $(-,-)$ & $(+,-)$ & $(-,+)$ & $(+,+)$ \\
\hline
$2^+_1$ & $1687.70$ & $629.26$ & $17.14$ & $140.57$ \\
$4^+_1$ & $2379.04$ & $876.18$ & $25.78$ & $198.76$ \\
$6^+_1$ & $2556.85$ & $920.91$ & $30.98$ & $214.70$ \\
$8^+_1$ & $2580.68$ & $899.62$ & $36.37$ & $218.34$ \\
$10^+_1$ & $2521.87$ & $841.70$ & $42.61$ & $215.48$ \\
\hline
\end{tabular}
\end{center}
\end{table}   

\subsection{SM results for multiple $SU(3)$ algebras:
$(sdg)^{(6_p,2_n)T=2}$ example}

In our second example, we have considered a system of 6 protons and 2 neutrons
in $\eta=4$ shell, i.e. $(sdg)^{6_p,2_n}$ system and carried out SM
calculations  using the four $H_Q$ Hamiltonians in the full SM space (dimension
in the $m$-scheme is $\sim 2 \times 10^7$)  using Antoine code. For this system,
the leading $SU(3)$ irrep (see Table II) is $(26,0)$ with $S=0$ and $T=2$. Then,
Eq. (\ref{eq.su310}) gives $E_{gs}=754$ and SM calculations for all four $H_Q$'s
is in agreement with this $SU(3)$ result. Also, in the SM results the excitation
energies of the yrast $J$  states or ground band members ($J=0,,2,4,6,\ldots$)
are seen to follow for all  the four $H_Q$'s the $3J(J+1)/4$ law as given by
$SU(3)$.  Thus, it is again verified by explicit SM calculations that all the
four $H_Q$'s give $SU(3)$ symmetry. The wavefunctions of the yrast $J$ states
are investigated by calculating $Q(J)$ and $B(E2)$'s for the ground band members
using $T^{E2}$ in Eq. (\ref{eq.e2qpqn}). In all the calculations,
$e^p_{eff}=1.5e$, $e_{eff}^n=0.5e$ and $b^2=A^{1/3}\,fm^2$ with $A=88$ are used.
Note that the ground $(26,0)$ irrep arises from the strong coupling of the
$(18,0)$ irrep for the $6$ protons (see the previous Section) and the $(8,0)$
irrep for the two neutrons. Therefore, formulas in Eq. (\ref{eq.e2formu}) will
apply for the states from $H_Q(-,-)$. Results from SM for the four $H_Q$'s are
given in Tables IV. It is easy to see that the results for $H_Q^{(-,-)}$ are in
complete agreement with the formulas in Eq. (\ref{eq.e2formu}). This is expected
as the proton and neutron parts of $T^{E2}$ in Eq. (\ref{eq.e2qpqn}) are 
generators of $SU^{(-,-)}(3)$ for protons and neutrons respectively. However,
the results from the other three $H_Q$'s are quite different as in the previous
$(sdg)^{6_p}$ example. Again, it is seen from Tables IV  that the results for
$Q(J)$ and $B(E2)$'s from $H_Q^{(+,-)}$ are closer to those from $H_Q^{(-,-)}$.
The $B(E2)$'s from $H_Q^{(-,+)}$ and $H_Q^{(+,+)}$ are much smaller in
magnitude. Moreover, $H_Q^{(-,-)}$ generates prolate shape  and $H_Q^{(+,+)}$
oblate as in the previous example. Finally, let us mention that we have also
carried out DSM calculations for this example and they are all in agreement with
SM results. 

\begin{table}
\begin{center}

\caption{Deformed shell model results for quadrupole moments  $Q(J)$ and $B(E2;
J \rightarrow J-2)$ values for the ground $K=0^+$ band members for a system of 6
protons and 6 neutrons (with$T=0$) in  $\eta=4$ shell. Results are given for the
four Hamiltonians in  Eq. (\ref{eq.su38}). In the table $(-,-)$ means we are
using the wavefunctions  obtained using $H_Q^{(-,-)}$ and similarly others.
Numbers in the brackets in the second column are exact $SU(3)$ results for 
$H_Q^{(-,-)}$. For other details see text.}

\begin{tabular}{ccccc}
\hline
$J$ & \multicolumn{4}{c}{$Q(J)\; efm^2$} \\
& $(-,-)$ & $(+,-)$ & $(-,+)$ & $(+,+)$ \\
\hline
$2^+_1$ & $-96.67 (-95.31)$ & $-65.95$ & $-4.63$ & $26.1$ \\
$4^+_1$ & $-123.04 (-123.12)$ & $-82.61$ & $-7.02$ & $33.41$ \\
$6^+_1$ & $-135.34 (-135.43)$ & $-88.65$ & $-9.66$ & $37.03$ \\
$8^+_1$ & $-142.45 (-142.58)$ & $-90.30$ & $-12.91$ & $39.24$ \\
$10^+_1$ & $-147.09 (-147.21)$ & $-89.60$  & $-16.86$ & $40.63$ \\
\hline
$J$ & \multicolumn{4}{c}{$B(E2; J \rightarrow J-2)\; e^2fm^4$} \\
& $(-,-)$ & $(+,-)$ & $(-,+)$ & $(+,+)$ \\
\hline
$2^+_1$ & $2273.96 (2276.93)$ & $1069.13$ &  $4.61$ & $164.89$ \\
$4^+_1$ & $3225.34 (3229.59)$ & $1503.04$ & $7.43$ & $233.99$ \\
$6^+_1$ & $3506.56 (3511.13)$ & $1608$ & $9.98$ & $254.61$ \\
$8^+_1$ & $3601.29 (3605.99)$ & $1613.12$ & $13.44$ & $261.78$ \\
$10^+_1$ & $3605.81 (3610.51)$ & $1565.64$ & $18.32$ & $262.44$ \\
\hline
\end{tabular}
\end{center}
\end{table}   

\subsection{DSM results for multiple $SU(3)$ algebras:
$(sdg)^{(6_p,6_n)T=0}$ example}

In our final example we have considered a system of 12 nucleons with $T=0$ in
$\eta=4$ shell, i.e. $(sdg)^{(6_p,6_n)T=0}$ system. Here the dimension in the
$m$-scheme in SM is $ \sim 10^{10}$ and therefore SM calculations are not
possible with our computational facilities. Thus, in this example DSM gives the
predictions for four $H_Q$'s and only for $H_Q^{(-,-)}$ we have exact $SU(3)$
results (they will be same as SM results if performed) from Section III.A.
Carrying out DSM calculations for this system,  it is found that the four
$H_Q$'s generate the same  HF sp spectrum as shown in Fig. 1. Using the lowest
intrinsic shown in Fig. 1, it is seen from the intrinsic quadrupole moments for
the four $H$'s that $H_Q^{(-,-)}$ generates prolate shape and $H_Q^{(+,+)}$
generates  oblate shape in agreement with SM. The ground state energy for the 
system is found to be $-1402.4$ for all four $H_Q$'s against the exact $SU(3)$
value $-1404$ giving less than $1\%$ deviation. Note that the $SU(3)$ irrep for
the ground band is $(36,0)$ and this generated by the irrep $(18,0)$ for the 6
protons and $(18,0)$ for the 6 neutrons. The energies of the yrast $J$ states
are also same for four $H_Q$'s and they are also within $1\%$ deviation from the
$3J(J+1)/4$ law. Turning to  $Q(J)$ and $B(E2)$'s, in the calculations used are
$e^p_{eff}=1.5e$, $e_{eff}^n=0.5e$ and $b^2=A^{1/3}\,fm^2$ with $A=92$. Note
that the ground $(36,0)$ irrep arises from the strong coupling of the $(18,0)$
irreps of the $6$ protons and the 6 neutrons. Therefore,
formulas in Eq. (\ref{eq.e2formu}) will apply for the states from $H_Q(-,-)$.
DSM results for $H_Q^{(-,-)}$, as shown in Table V are in complete agreement
with the formulas in Eq. (\ref{eq.e2formu}) as expected.  However, the results
from the other three $H_Q$'s are quite different as in the previous
$(sdg)^{6_p}$ and $(sdg)^{6p,2n}$ examples. Again, it is seen from Tables V that
the results for $Q(J)$ and $B(E2)$'s from $H_Q^{(+,-)}$ are closer to those from
$H_Q^{(-,-)}$. The $B(E2)$'s from $H_Q^{(-,+)}$ and $H_Q^{(+,+)}$ are much
smaller in magnitude. Moreover, $H_Q^{(-,-)}$ generates prolate shape  and
$H_Q^{(+,+)}$ oblate as in the previous examples. Thus, the results in Tables
III-V are generic results for the four $H_Q$'s. 

\section{Multiple $SU(3)$ algebras in interacting boson model}

In the interacting boson models with $sd$ ($\ell=0,2$) or $sdg$ ($\ell=0,2,4$)
bosons (and their appropriate generalizations to $pf$, $sdgi$ etc.), the eight 
operators $(L^1_q , Q^2_q(\baa))$ are
\be
\barr{l}
L^1_q = \dis\sum_\ell \dis\sqrt{\dis\frac{\ell(\ell+1)(2\ell+1)}{3}}
\l(b^\dagger_\ell \tilde{b}_\ell\r)^1_q \;,\\
Q^2_q(\baa)= -(2\eta+3)\,\dis\sum_\ell
\dis\sqrt{\dis\frac{\ell(\ell+1)(2\ell+1)}{5(2\ell+3)(2\ell-1)}}
\l(b^\dagger_\ell \tilde{b}_\ell\r)^2_q \\
+ \dis\sum_{\ell <
\eta}\,\alpha_{\ell,\ell+2}\;\dis\sqrt{\dis\frac{6(\ell+1)(\ell+2)(\eta-\ell)
(\eta+\ell+3)}{5(2\ell+3)}}\l[\l(b^\dagger_\ell \tilde{b}_{\ell+2}\r)^2_q +
\l(b^\dagger_{\ell+2} \tilde{b}_{\ell}\r)^2_q \r]\;;
\alpha_{\ell,\ell+2}=\pm 1\;.
\earr \label{eq.apb1p}
\ee
Note that $b^\dagger$ and $b$ are boson creation and annihilation operators and
$\tilde{b}_{\ell m} = (-1)^{\ell-m}b_{\ell -m}$. Again,  after some tedious 
angular momentum algebra, it is easy to prove that for all choices of 
$\alpha_{\ell,\ell+2} = \pm 1$, Eq. (\ref{eq.su36}) is valid and therefore
giving a $SU(3)$ algebra for each choice of the $\alpha$'s. With
$\alpha_{\ell,\ell+2}$ taking $+1$ or $-1$ value, for a given $\eta$ there will
be $2^{[\eta/2]}$ number of $SU(3)$ algebras in IBM's just as in SM. It is
important to stress that  $\alpha_{\ell,\ell+1}=+1$ for all $\ell$ values is the
standard choice in $sd$IBM and $sdg$IBM.  As an example, in $sd$IBM with
$\eta=2$, the $(L^1_q,Q^2_q)$ operators generating multiple $SU(3)$ algebra are,
\be
\barr{l}
L^1_q = \dis\sqrt{10}  \l(d^\dagger \tilde{d}\r)^1_q \;,\\
Q^2_q(\alpha_{sd}) = \dis\sqrt{2}\,\l[ -
\dis\frac{\dis\sqrt{7}}{2} \l(d^\dagger \tilde{d}\r)^2_q 
+ \alpha_{sd} \l(s^\dagger \tilde{d} + d^\dagger
\tilde{s}\r)^2_q\r]\;;\;\;\alpha_{sd}=\pm 1\;.
\earr \label{apb3}
\ee
giving two $SU^{\baa}(3)$ algebras. In $sd$IBM they are discussed in the context
of  quantum phase transitions (QPT) \cite{RMP}. The $\alpha_{sd}=+1$ and $-1$
generate prolate and oblate shapes respectively as discussed ahead.  In $sdg$IBM
with $\eta=4$ there will be four $SU^{\baa}(3)$ algebras generated by,
\be
\barr{l}
L^1_\mu = \dis\sqrt{10}  \l(d^\dagger \tilde{d}\r)^1_\mu + 2\dis\sqrt{15}  
\l(g^\dagger \tilde{g}\r)^1_\mu\;,\\
Q^2_\mu(\alpha_{sd},\alpha_{dg}) = 
\dis\sqrt{\dis\frac{3}{4}}\l\{ - 11\dis\sqrt{\dis\frac{2}{
21}}\,(d^{\dg} \tilde{d})^2_\mu -2 \dis\sqrt{\dis\frac{33}{7}}\, (g^{\dg} 
\tilde{g})^2_\mu \r. \\
\l. + \alpha_{sd}\,4 \dis\sqrt{\dis\frac{7}{15}} \;\left(s^{\dg}\tilde{d} +
d^{\dg}\tilde{s} \right)^2_\mu + \alpha_{dg}\,\dis\frac{36} {\dis\sqrt{105}}
\;\left(d^{\dg} \tilde{g} + g^{\dg} \tilde {d} \right)^2_\mu \r\} \;,
\earr \label{eq.apb4}
\ee
with$\alpha_{sd}=\pm 1$ and $\alpha_{dg}=\pm 1$. 

\begin{figure}[!tbh]
\begin{center}
\includegraphics[width=0.7\linewidth]{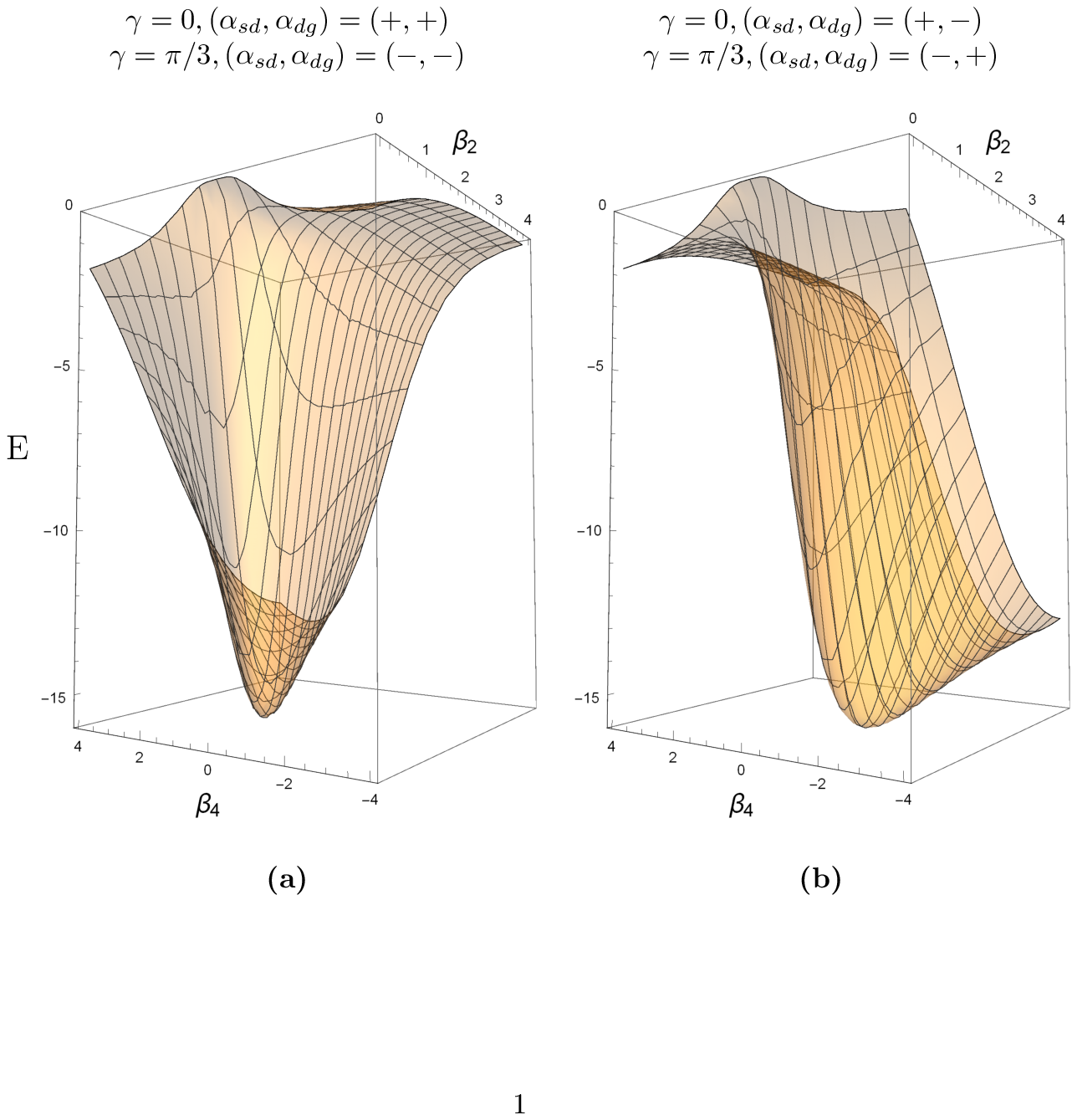}
\end{center}
\vskip -0.25cm
\caption{ Energy functional $E=E_{SU_{sdg}(3)}/N^2$ as a function of
$\beta_2$ and $\beta_4$. (a) plot for $(\alpha_{sd}=1, \alpha_{dg}=1)$ with
$\gamma=0^\circ$ and $(\alpha_{sd}=-1, \alpha_{dg}=-1)$ with $\gamma=60^\circ$.
Note that the energy functional is same for both of these choices as can be seen
from Eq.(\ref{eq.apb7}). (b) Same as (a) but for $(\alpha_{sd}=1,
\alpha_{dg}=-1)$  with $\gamma=0^\circ$ and $(\alpha_{sd}=-1, \alpha_{dg}=+1)$
with  $\gamma=60^\circ$.}
\label{csfig}
\end{figure}
 
\subsection{Geometry of multiple $SU(3)$ algebras in $sd$IBM and $sdg$IBM}

In order to have some insight into the multiple $SU(3)$ algebras in IBM,  let us
examine the geometric shapes generated by them using coherent states. Starting
with $sd$IBM, the coherent state is
\be
\l.\l| N;\beta_{2};\gamma \r.\ran   =  \l[ N! \l(1+\beta_{2}^{2} \r)^{N} \r]^{
-1/2} \l\{ s^{\dagger}_{0} +\beta_{2}
\l[ \cos \gamma\,\, d^{\dagger}_{0} + \sqrt {\frac{1}{2}} \sin \gamma 
\l(d^{\dagger}_{2}+d^{\dagger}_{-2} \r)\r] \r\}^N \;,
\label{eq.sdcoher}
\ee
where $\beta_2 \geq 0$ and $0^\circ  \leq \gamma \leq 60^\circ$. Now, let us 
consider the $SU(3)$ Hamiltonian 
\be
H^{\alpha_{sd}}_{SU_{sd}(3)}=-Q^2(\alpha_{sd}) \cdot Q^2(\alpha_{sd})
\label{eq.xx1}
\ee
and $-Q^2(\alpha_{sd}) \cdot Q^2(\alpha_{sd}) = -C_2(SU^{\alpha_{sd}}(3)) + 
\frac{3}{4} L \cdot L$. It is
important to note that $H_{SU_{sd}(3)}$ generates the same spectrum for
the two choices of $\alpha_{sd}$. In the  $N \rightarrow \infty$ limit, the
coherent state expectation value of  $H_{SU_{sd}(3)}$ is given by
\be
\barr{l} 
E_{SU^{\alpha_{sd}}_{sd}(3)}\l(N;\beta_2, \gamma\r) = \lan N;\beta_{2},
\gamma \mid -Q^2(\alpha_{sd}) \cdot Q^2(\alpha_{sd}) \mid N;\beta_{2},
\gamma \ran \\
=-\dis\frac{2N^2}{\l(1+\beta_2^2\r)^2}\;\l[
4\beta_2^2 + \dis\frac{\beta^4_2}{2} + 2\dis\sqrt{2}\,\alpha_{sd}\,
\beta_2^3\,\cos3\gamma\r]\;.
\earr \label{eq.sdenergy}
\ee
Minimizing the $SU(3)$ energy functional $E_{SU_{sd}(3)}\l(N;\beta_2, \gamma\r)$
gives the equilibrium solutions $(\beta_2^0 , \gamma^0)$ to be $\beta_2^0 =
\sqrt{2}$ and $\gamma^0=0^\circ$ for $\alpha_{sd}=+1$ and $\gamma^0= 60^\circ$
for $\alpha_{sd}=-1$. Also, for both situations the equilibrium energy is
$-4N^2$ and this is same as the large $N$ eigenvalue of $-C_2(SU(3))$ in the
h.w. $(2N,0)$ irrep [also for the lowest weight $(0,2N)$ irrep].  Note that the
eigenvalue of $C_2(SU(3))$ in a $SU(3)$ irrep $(\lambda \mu)$ is simply
$\lambda^2 + \mu^2 + \lambda \mu + 3(\lambda + \mu)$. Also, the formula in Eq.
(\ref{eq.sdenergy}) is good in the limit $N \rightarrow \infty$ and in this
limit $L \cdot L$ will not contribute as only terms of the order of $N^2$ will
survive. Thus, $\alpha_{sd}=\pm 1$ will give prolate and oblate solutions and 
these results for $sd$IBM are well known \cite{Iac-87,RMP}.

First non-trivial situation happens with $sdg$IBM and for this we will consider 
the three parameter coherent state used in \cite{KY,Piet} in terms of 
$(\beta_2,\beta_4,\gamma)$ parameters for a $N$ boson system,
\be
\barr{l}
\vspace{3mm}
\l| N;\beta_{2};\beta_{4},\gamma \ran   =  \left[ N! \left(
1+\beta_{2}^{2}+\beta_{4}^{2} \right)^{N} \right]^{-1/2}
\left\{ s^{\dagger}_{0} +\beta_{2}
\left[ \cos \gamma\,\, d^{\dagger}_{0} +  \right. \right. \\
\left. \left. \sqrt {\frac {1} {2}} \sin \gamma \left(
d^{\dagger}_{2}+d^{\dagger}_{-2} \right)
\right] + \frac{1}{6}\beta_{4} \left[ \left(5\cos^{2}\gamma + 1\right)
g^{\dagger}_{0} \right. \right. \\
\left. \left. + \sqrt {\frac {15} {2}} \sin 2\gamma\,\left(
g^{\dagger}_{2}+g^{\dagger}_{-2}\right)
+ \sqrt {\frac {35} {2}} \sin^{2}\gamma\, \left(
g^{\dagger}_{4}+g^{\dagger}_{-4}\right) \right] \right\}^{N}\,\l|  0 \ran\;.
\earr \label{eq.apb5}
\ee
Note that  $\beta _{2} \ge  0$, $-\infty  \le  \beta _{4} \le +\infty$ and
$0^{\circ } \le  \gamma  \le  60^{\circ }$ respectively. Using the results given
\cite{KY,KY-rev}, the $SU(3)$ energy functional is given by 
\be
\barr{l} 
E_{SU^{\baa}_{sdg}(3)}\l(N;\beta_2,\beta_4,\gamma\r) = 
\lan N;\beta_{2};\beta_{4},
\gamma \mid -Q^2(\baa) \cdot Q^2(\baa) \mid N;\beta_{2};\beta_{4},\gamma \ran \\
=\dis\frac {-3N^2}{4\l(1+\beta_2^2+\beta_4^2\r)^2}\;\l[
\dis\frac{448}{15}\,\alpha_{sd}^2\,\beta_2^2 + 
\dis\frac{384\dis\sqrt{14}}{35}\,\alpha_{sd} \alpha_{dg}\,\beta_2^2\;\beta_4 
\r. \\ 
+ \dis\frac{352\dis\sqrt{35}}{105}\,\alpha_{sd}\,
\beta_2^3\,\cos{3\gamma} + \dis\frac{64\dis\sqrt{35}}{21}\,\alpha_{sd}
\beta_2\,\beta_4^2\,\cos{3\gamma} +
\dis\frac{3456}{245}\,\alpha_{dg}^2\,\beta_2^2\,\beta_4^2 \\  
+ \dis\frac{1056\dis\sqrt{10}}{245}\,\alpha_{dg}\,
\beta_2^3\,\beta_4\,\cos{3\gamma} + \dis\frac{484}{147}\,\beta_2^4 +
\dis\frac{192\dis\sqrt{10}}{49}\,\alpha_{dg}\,
\beta_2\,\beta_4^3\,\cos{3\gamma} \\ 
+ \l. \dis\frac{880}{441}\,
\l(4-{\cos}^23\gamma\r)\,\beta_2^2\beta_4^2 + \dis\frac{400}{1323}\,
\l(16-7{\cos}^23\gamma\r)\,\beta_4^4 \r] \;.
\earr \label{eq.apb7}
\ee
Note that $\baa=(\alpha_{sd} , \alpha_{dg})$. Minimizing $E_{SU_{sdg}(3)}
\l(N;\beta_2,\beta_4,\gamma\r)$ with respect to $\beta_2$, $\beta_4$ and
$\gamma$ will give the equilibrium (ground state) shape parameters ($\beta_2^0$,
$\beta_4^0$, $\gamma^0)$ and the corresponding equilibrium energy
$E^0_{SU_{sdg}(3)}$. Results are given in Table \ref{shape}.  
\begin{table*}

\caption{Equilibrium shapes for the four $SU(3)$ algebras in $sdg$IBM. For
$(\alpha_{sd},\alpha_{dg}) = (-1,+1)$ and $(-1,-1)$, shown are the $\beta_2^0$
and $\beta_4^0$ values for both $\gamma^0=0^{\circ }$ and $60^{\circ }$ and they
are equivalent.}

\begin{tabular}{cccccl}
\hline 
\hline
$\alpha_{sd}$ & $\alpha_{dg}$ & $\beta_2^0$ & $\beta_4^0$ & $\gamma^0$ &
$E^0_{SU_{sdg}(3)}$ \\
\hline
\hline
+1 & +1 & $\sqrt{20/7}$ & $\sqrt{8/7}$ & $0^{\circ }$ & $-16N^2$ \\
+1 & -1 & $\sqrt{20/7}$ & $-\sqrt{8/7}$ & $0^{\circ }$ & $-16N^2$ \\
-1 & +1 & $\sqrt{20/7}$ & $-\sqrt{8/7}$ & $60^{\circ }$ & $-16N^2$ \\
& & $-\sqrt{20/7}$ & $-\sqrt{8/7}$ & $0^{\circ }$ & $-16N^2$ \\
-1 & -1 & $\sqrt{20/7}$ & $\sqrt{8/7}$ & $60^{\circ }$ & $-16N^2$ \\
& & $-\sqrt{20/7}$ & $\sqrt{8/7}$ & $0^{\circ }$ & $-16N^2$ \\
\hline\hline
\end{tabular}
\label{shape}
\end{table*}
As seen from the Table \ref{shape}, the four values of
$(\alpha_{sd},\alpha_{dg})$  generate four combinations of
$(\beta^0_2,\beta^0_4,\gamma^0)$. These can be easily understood from the
symmetries under $\beta_2 \rightarrow -\beta_2$, $\beta_4 \rightarrow -\beta_4$
and $\gamma=0^\circ \rightarrow 60^\circ$. We have for example $E(\beta_2,
\beta_4, \gamma; \alpha_{sd}=1, \alpha_{dg}=1) = E(\beta_2, -\beta_4, \gamma;
\alpha_{sd}=1, \alpha_{dg}=-1)$,  $E(\beta_2, \beta_4, \gamma; \alpha_{sd}=1,
\alpha_{dg}=1) = E(\beta_2, \beta_4, \gamma+60^\circ ; \alpha_{sd}=-1,
\alpha_{dg}=-1) = E(-\beta_2, -\beta_4, \gamma; \alpha_{sd}=-1, \alpha_{dg}=1) =
E(-\beta_2, \beta_4, \gamma; \alpha_{sd}=-1, \alpha_{dg}=-1)$. These also show
that the solutions with $\gamma=60^{\circ}$ can be changed to $\gamma=0^{\circ}$
with $\beta_2 \rightarrow -\beta_2$ as given in Table \ref{shape}. More
importantly, for all the four  solutions, the $E^0_{SU_{sdg}(3)}=-16N^2$. This
energy value  is same as the large $N$ eigenvalue of $-C_2(SU(3))$ in $(4N,0)$
irrep. This then implies that the internal structure of the $(4N,0)$ irrep is
different for the four solutions as discussed ahead. The energy functional is
shown in Fig. \ref{csfig} as a function of $\beta_2$ and $\beta_4$  for
$\gamma=0^\circ$ and $60^\circ$ for the four choices of $(\alpha_{sd} ,
\alpha_{gd})$. 

\subsection{Large $N$ results for quadrupole moments and $B(E2)$'s}

For further understanding of the four solutions for $SU_{sdg}(3)$, we have
examined quadrupole moments and $B(E2)$ values in the ground $K=0$ band 
generated by the four solutions in Table \ref{shape}. Note that the intrinsic
state structure for the $K=0$ ground band is 
\be
\l|N;K=0\ran = (N!)^{-1/2} \l(x_0 s^\dagger_0 + x_2
d^\dagger_0 +x_4 g^\dagger_0\r)^N \l|0\ran \;
\label{eq.ground}
\ee
where $x_0=\sqrt{1/5}$, $x_2=\beta_2^0/\sqrt{5}$ and $x_4=\beta_4^0/\sqrt{5}$ 
with $\gamma^0=0^{\circ}$. It is easy to construct the angular momentum
projected states $\l|N;K=0,L,M\ran$ and calculate quadrupole moments $Q(L)$ and
$B(E2; L \rightarrow L-2)$ for the ground band. The formulation for these is
given in detail in \cite{Kuyu} and valid to order $1/N^2$ where $N$ is the boson
number.  Then we have,
\be 
\barr{l}
Q(L)=\lan LL \mid Q^2_0 \mid LL\ran = \dis\frac{\lan LL\;20 \mid LL\ran}{
\dis\sqrt{2L+1}}\;\lan L \mid\mid Q^2 \mid\mid L\ran\;,\\
B(E2;L \rightarrow L-2)= \dis\frac{5}{16\pi}\;\f{\l|\lan L-2 \mid\mid Q^2 
\mid\mid L\ran\r|^2}{(2L+1)}\;;\\
\\
\lan N;K=0,L_f \mid\mid Q^2 \mid\mid N;K=0,L_i\ran =
\l[N\dis\sqrt{(2L_i+1)}\r] \lan L_i 0\;\;20 \mid L_f,0\ran \;\times \\
\l[B_{00} +\frac{1}{N}\l(B_{00} - \dis\frac{B_{10}-3B_{00}}{a}\r) 
-\dis\frac{L_i(L_f+1)}{aN^2}\l\{B_{00}
+\dis\frac{F_1}{4a}\;\delta_{L_f, L_i} \r.\r. \\
\l.\l. - \dis\frac{F_2}{12a}\;\delta_{L_f, L_i+2}
\r\}\r]\;; \;\;\;\;L_f=L_i\;\;\mbox{or}\;\;L_f=L_i+2\\
B_{mn}=\dis\sum_{\ell^\pr , \ell}\;\l[\ell^\pr(\ell^\pr+1)\r]^m 
\l[\ell (\ell+1)\r]^n \lan \ell^\pr
0\;\ell 0 \mid 20\ran \; t_{\ell^\pr,\ell}\;x_{\ell^\pr 0} x_{\ell 0}\;, \\
F_1=B_{20}-B_{11}-10B_{10}+12B_{00},\;\;F_2=B_{20}-B_{11}
+6B_{10}-12B_{00}, \\
a=\dis\sum_{\ell} \ell (\ell+1) \l(x_{\ell}\r)^2,\;\;\;\ell=0,\;2,\;4\;.
\earr \label{eq.qbe2}
\ee
In Eq. (\ref{eq.qbe2}), the $t_{\ell^\pr , \ell}$ are the coefficients in
the $E2$ transition operator and they are chosen as, 
\be
T^{E2} = \dis\sum_{\ell^\pr , \ell} t_{\ell^\pr \ell} \l(b^\dagger_{\ell^\pr}
\tilde{b}_\ell\r)^2_q = Q^2_q(\alpha_{sd}=+1,\alpha_{dg}=+1) \;.
\label{eq.e2op}
\ee
See Eq. (\ref{eq.apb4}) for $Q^2_q(\alpha_{sd}=+1,\alpha_{dg}=+1)$. Using the
$T^{E2}$, the solutions in Table \ref{shape} and Eq. (\ref{eq.qbe2}), results
are obtained for $Q(2^+_1)$, $Q(4^+_1)$, $B(E2; 2^+_1 \rightarrow 0^+_1)$ and
$B(E2; 4^+_1 \rightarrow 2^+_1)$ for a 10 boson system and the results are given
in Table \ref{Qbe2}. It is seen that the $SU^{(+,+)}(3)$ and $SU^{(+,-)}(3)$ are
closer generating prolate shape and $SU^{(-,-)}(3)$ generating oblate shape.
The $SU^{(-,+)}(3)$ though generates prolate shape, the quadrupole moments are
very small. Thus, $sdg$IBM substantiates the general structures observed in $sdg$
shell model examples presented in Section III.

\begin{table*}
\caption{Quadrupole moments and $B(E2)$ values for low-lying states in the
ground band for a 10 boson system generated by the four $SU(3)$ algebras in 
$sdg$IBM. Note that $T^{E2}$ Eq. (\ref{eq.e2op}) is unit-less and therefore 
$Q(L)$ and $B(E2)$'s in the table are unit-less.}
\begin{tabular}{cccccl}
\hline 
\hline
$\alpha_{sd}$ & $\alpha_{dg}$ & $Q(2^+_1)$ & $Q(4^+_1)$ & $B(E2; 2^+_1 
\rightarrow 0^+_1)$ & $B(E2; 4^+_1 \rightarrow 2^+_1)$  \\
\hline
\hline
+1 & +1 & $-13.69$ & $-17.43$ & $45.68$ & $64.87$ \\
+1 & -1 & $-6.15$ & $-7.89$ & $9.15$ & $12.61$ \\
-1 & +1 & $-2.16$ & $-2.98$  & $1.05$ & $1.59$ \\
-1 & -1 & $5.38$ & $6.55$ & $7.33$ & $10.51$ \\
\hline\hline
\end{tabular}
\label{Qbe2}
\end{table*}

\section{Conclusions}

Multiple $SU(3)$ algebras appear in both shell model and interacting boson model
spaces and they open a new paradigm in the applications of $SU(3)$ symmetry in
nuclei. In the first detailed attempt made in this paper, using three $(sdg)$
space examples in SM, we showed that the four $SU(3)$ algebras in this space
exhibit quite different properties with regard to quadrupole collectivity as
brought out by the quadrupole moments $Q(J)$ and $B(E2)$'s in the ground $K=0$ band in
even-even systems (see Tables III-V). The SM and DSM calculations are restricted
to the examples with the leading $SU(3)$ irrep of the type $(\la 0)$. The
prolate, oblate and intermediate  structures from the four $SU(3)$ algebras
found using SM and DSM is further  substantiated by coherent state analysis and
asymptotic formulas for quadrupole moments and $B(E2)$'s in the ground band in
$sdg$IBM. Also, the results from $Q(J)$ and $B(E2)$'s for the four $SU(3)$
algebras are consistent with the correlation coefficients between the four
different $Q.Q$ operators in the $sdg$ space of SM. Results in Tables III-V and
VII may be useful in finding empirical examples  for multiple $SU(3)$ algebras
in $sdg$ and larger SM spaces and in $sdg$IBM.

Going beyond the present investigations, in future the structure of the
low-lying  $\gamma$ (also $\beta$) band generated by the multiple $SU(3)$
algebras will be investigated using SM and DSM. Here, we need to deal with the
$SU(3)$ integrity basis operators that are 3 and 4-body, as the leading  $SU(3)$
irrep in general will be of the type $(\la \mu)$ with $\mu \neq 0$ \cite{JPD-1}.
For example, as seen from Table II, for $8$ nucleons with $T=0$ the leading
$SU(3)$ irrep is $(24,4)$. Let us add that the method for dealing with 3-body
operators in DSM was described in \cite{KS}. In addition, applications of the
$H_Q$'s in Eq. (\ref{eq.su38}) to quantum phase transitions (QPT) may give new
insights. For example, using $H=\sum_{\baa} c_\baa Q^2(\baa) \cdot Q^2(\baa)$
and varying the parameters $c_\baa$, it is possible to study QPT; for a similar
study using multiple pairing algebras in SM and IBM see \cite{Ko-BJP}. Also
studies using $Q^2(\baa , p) \cdot Q^2(\baa^\prime , n)$ with $\baa \neq
\baa^\prime$ and $p$ ($n$) denoting protons (neutrons) will be of interest;
results of such a study in $sd$IBM are known \cite{DB-82}. In $sdg$IBM a more
general CS in terms of  $(\beta_2, \beta_4, \gamma ,\gamma_4 , \delta_4)$ given
in  \cite{Ydd,Piet-cs,Roz} may prove to be important in understanding further
the four $SU(3)$ algebras in this model. Also, it is possible to examine the
properties of $\beta$ and $\gamma$ bands in this model using the results in
\cite{Kuyu,Kuyu-2}. All these will be addressed and the results will be reported
in a future publication.

\section*{Acknowledgments} 

RS is thankful to SERB of DST of Government of India for financial support.

\ed